\documentclass[
superscriptaddress,
 amsmath,amssymb,
 aps,
prl,
onecolumn,
]{revtex4-2}

\usepackage{xpatch}
\makeatletter
\xpatchcmd\@collaboration@present{(}{\medskip}{}{}
\xpatchcmd\@collaboration@present{)}{}{}{}
\makeatother

\newcommand{\thor}{$^{229}$Th}

\newcommand{\thornitr}{$\rm Th(NO_3)_4$}
\newcommand{\lisaf}{LiSrAlF$_6$}
\newcommand{\sapph}{$\textrm{Al}_2\textrm{O}_3$}
\newcommand{\mgf}{MgF$_2$}
\newcommand{\thox}{$^{229}$ThO$_2$}
\newcommand{\regthf}{ThF$_4$}
\newcommand{\thf}{\thor F$_4$}

\usepackage[dvipsnames]{xcolor}
\definecolor{ricky}{cmyk}{0, 0.7808, 0.4429, 0.1412}
\definecolor{jack}{cmyk}{0,0.49,0.59,0.32}
\definecolor{chuankun}{cmyk}{0,1,0,0}

\usepackage{filecontents}
\usepackage{siunitx}
\usepackage{braket}
\usepackage{graphicx}
\usepackage[colorlinks=true, allcolors=blue]{hyperref}
\usepackage{gensymb}
\usepackage{upgreek}

\usepackage{float}  
\usepackage{xcolor}

\usepackage{braket}
\usepackage{graphicx}

\usepackage{bibunits}
\defaultbibliographystyle{apsrev4-2}
\usepackage{etoolbox}

\makeatletter
\newcommand*{\newbibstartnumber}[1]{%
  \apptocmd{\thebibliography}{%
    \global\c@NAT@ctr #1\relax
    \addtocounter{NAT@ctr}{-1}%
  }{}{}%
}
\makeatother

\usepackage[colorlinks=true, allcolors=blue]{hyperref}
\usepackage[version=3]{mhchem}

\begin{document}

\title{\thf{} thin films for solid-state nuclear clocks}

\author{Chuankun Zhang}
\author{Lars von der Wense}
\author{Jack F. Doyle}
\author{Jacob S. Higgins}
\author{Tian Ooi}
\author{Hans U. Friebel}
\author{Jun Ye}
\affiliation{JILA, NIST and University of Colorado, Department of Physics, University of Colorado, Boulder, CO 80309}

\makeatletter
    \let\tmpaffiliation\affiliation
    \let\tmpauthor\author
    \let\tmpabstract\frontmatter@abstract@produce
    \let\frontmatter@abstract@produce\relax
    \let\frontmatter@finalspace\relax
    \maketitle
    
    \def\frontmatter@finalspace{\addvspace{18\p@}}
    \let\maketitle\frontmatter@maketitle
    \let\affiliation\tmpaffiliation
    \let\author\tmpauthor
    \let\frontmatter@abstract@produce\tmpabstract
    \let\frontmatter@title@produce\relax
    \makeatother

\author{R. Elwell}
\affiliation{Department of Physics and Astronomy, University of California, Los Angeles, CA 90095, USA}
\author{J. E. S. Terhune}
\affiliation{Department of Physics and Astronomy, University of California, Los Angeles, CA 90095, USA}
\author{H. W. T. Morgan}
\affiliation{Department of Chemistry and Biochemistry, University of California, Los Angeles, Los Angeles, CA 90095, USA}
\affiliation{Department of Chemistry, University of Manchester, Oxford Road, Manchester M13 9PL, UK}
\author{A. N. Alexandrova}
\affiliation{Department of Chemistry and Biochemistry, University of California, Los Angeles, Los Angeles, CA 90095, USA}
\author{H. B. Tran Tan}
\affiliation{Department of Physics, University of Nevada, Reno, Nevada 89557, USA}
\affiliation{Los Alamos National Laboratory, P.O. Box 1663, Los Alamos, New Mexico 87545, USA} 
\author{Andrei Derevianko}
\affiliation{Department of Physics, University of Nevada, Reno, Nevada 89557, USA}
\author{Eric R. Hudson}
\affiliation{Department of Physics and Astronomy, University of California, Los Angeles, CA 90095, USA}
\affiliation{Challenge Institute for Quantum Computation, University of California Los Angeles, Los Angeles, CA, USA}
\affiliation{Center for Quantum Science and Engineering, University of California Los Angeles, Los Angeles, CA, USA}

\date{\today} 



\maketitle

\begin{bibunit}
\textbf{After nearly fifty years of searching, the vacuum ultraviolet \thor{} nuclear isomeric transition has recently been directly laser excited  ~\cite{Tiedau2024,Elwell2024} and measured with high spectroscopic precision~\cite{Zhang2024}.
Nuclear clocks based on this transition are expected to be more robust~\cite{Hudson2008, Rellergert2010} than and may outperform~\cite{PeikTamm2003,Campbell2012} current optical atomic clocks. They also promise sensitive tests for new physics beyond the standard model~\cite{Rellergert2010,Flambaum2006,Litvinova2009}. 
In light of these important advances and applications, a dramatic increase in the need for \thor{} spectroscopy targets in a variety of platforms is anticipated.
However, the growth and handling of high-concentration \thor{}-doped crystals~\cite{Rellergert2010} used in previous measurements~\cite{Jeet2015,Tiedau2024,Elwell2024,Zhang2024} are challenging due to the scarcity and radioactivity of the \thor{} material.
Here, we demonstrate a potentially scalable solution to these problems by demonstrating laser excitation of the nuclear transition in \thor F$_4$ thin films grown with a physical vapor deposition process, consuming only micrograms of \thor{} material.
The \thf{} thin films are intrinsically compatible with photonics platforms and nanofabrication tools for integration with laser sources and detectors, paving the way for an integrated and field-deployable solid-state nuclear clock with radioactivity up to three orders of magnitude smaller than typical \thor-doped crystals~\cite{Jeet2015, Tiedau2024, Elwell2024, Zhang2024}.
The high nuclear emitter density in \thf{} also potentially enables quantum optics studies in a new regime.
Finally, we describe the operation and present the estimation of the performance of a nuclear clock based on a defect-free ThF$_4$ crystal.}
\\

The \thor{} nucleus has been at the focus of intense scientific investigation for several decades, as it possesses an extraordinarily low-energy isomeric nuclear transition at about 8.4 eV. 
This feature permits precision laser spectroscopy and has led to proposals for multiple important applications~\cite{Tkalya1996, Rellergert2010, Tkalya2011}, including the development of a nuclear optical clock~\cite{vonderwense2020review, beeks2021, peik2021review}. 
Of particular interest, a solid-state nuclear clock~\cite{Hudson2008, Rellergert2010, Kazakov2012} can be constructed by tightly confining the nuclei in a high bandgap crystalline lattice.
This confinement enables laser-based, recoil-free Mössbauer spectroscopy without sophisticated laser cooling and trapping techniques. 
Due to small nuclear electromagnetic moments, the nuclear clock transition frequency remains relatively insensitive to external perturbations in the crystalline host. Probing macroscopic amounts ($>10^{12}$) of \thor{} atoms in a solid promises excellent counting statistics and good clock stability.
An optical clock based on such a solid-state \thor{} sample is thus highly desirable for field applications due to its potential robustness and simplicity.

Recently, rapid progress has been made in this field. 
Direct laser excitation and measurement of the nuclear clock transition~\cite{Tiedau2024, Elwell2024} were demonstrated, and subsequently, 
high-precision spectroscopy referenced to the $^{87}$Sr clock~\cite{Zhang2024} was performed via a VUV frequency comb to determine the transition frequency to the $10^{-12}$ level. 
These experiments lay the foundation for a solid-state nuclear clock based on \thor{} doped into high bandgap crystals. 
However, the scarcity and the radioactivity of \thor{} severely complicate the crystal growing process. 
\thor{} is a controlled material that does not naturally exist, cannot be produced conveniently, and is already in high demand for medical research~\cite{hogle2016}. 
While the estimated world stock of \thor{} with reasonable isotopic purity is $\sim 40$ g~\cite{forsberg1999}, most of it is mixed with other chemical elements like $^{233}$U and is thus not readily available. 
The amount of \thor{} available to researchers is typically limited to at most a few milligrams, necessitating challenging miniaturized crystal growing techniques~\cite{Jeet2015, JeetThesis2018, Beeks2023} to obtain crystals with sufficient \thor{} dopant density. 
Further, given that cutting and polishing techniques require crystals with dimensions on the millimeter scale, these crystals typically have a radioactivity of $\gtrsim 10$~kBq. 
This level of radioactivity requires appropriate radiation safety precautions and limits the proliferation of nuclear clocks.

An alternative to crystal growth is physical vapor deposition (PVD), where a desired material is evaporated from a hot crucible and subsequently condensed onto a substrate. 
PVD is frequently used to produce radioactive targets~\cite{sletten1972, adair1979, glover1981, maier1981, maier1991, greene1993}, mainly for accelerator facilities. 
As $^{232}$\regthf{} thin films with the naturally abundant isotope are routinely used in optics, PVD of $^{232}$\regthf{} is a mature technology~\cite{Baumeister1973}.
However, PVD of \thf{} is much less developed with, to the best of our knowledge, only one brief description in the literature~\cite{glover1981}. 
Here, we report the fabrication of thin film \thf{} targets  (30-100~nm thickness) by miniaturized radioactive PVD on various substrates such as \sapph{} and \mgf. 
We fabricate targets as small as \SI{50}{\micro\meter} diameter using only micrograms of \thor{} material, orders of magnitude smaller than that needed for a single crystal growth batch~\cite{Rellergert2010b, JeetThesis2018, Beeks2023}. 
By reducing the target area without sacrificing the target optical density, the \SI{50}{\micro\meter} target has three orders of magnitude smaller radioactivity than typical \thor-doped crystals, corresponding to a much reduced radioactive hazard~\cite{iaea2018} and thus vastly relaxed safety measures.

While ThF$_4$ was originally assumed to have a band gap below the 8.4 eV \thor{} isomer energy~\cite{Jain2013} and therefore not considered as a suitable host for a solid-state nuclear clock~\cite{Rellergert2010b, JeetThesis2018}, recent work~\cite{Ellis2014,Gouder2019} indicates that the band gap is roughly 10~eV. 
Very recently, photo- and radio-luminescence of $^{232}$\regthf{} thin films were measured~\cite{osipenko2024}, highlighting a low fluorescence background that is beneficial for nuclear clock operation.
This suggests the possibility of observing the radiative decay of the \thor{} nuclear clock isomer in \thf. 
To confirm this, we perform nuclear laser spectroscopy on PVD \thf{} films using a vacuum ultraviolet (VUV) laser system.
We observe a single spectroscopic line at $2020406.8(4)_\textrm{stat}(30)_\textrm{sys}$~GHz for a \thf{} target on the \mgf{} substrate and at $2020409.1(7)_\textrm{stat}(30)_\textrm{sys}$~GHz with the \sapph{} substrate, which we assign to the \thor\ nuclear isomeric transition based on the prior crystal results~\cite{Tiedau2024,Elwell2024}; here and throughout numbers in parentheses denote 68\% confidence intervals.
The measured lifetimes in these samples, $150(15)_{\textrm{stat}}(5)_{\textrm{sys}}$~s on \sapph\ and $153(9)_{\textrm{stat}}(7)_{\textrm{sys}}$~s on \mgf, are identical within error, and are substantially shorter than that observed in \thor{}:CaF$_2$ ($\tau = 641(4)$~s~\cite{Zhang2024}) and \thor{}:\lisaf{} ($\tau =  568(13)_{\textrm{stat}}(20)_{\textrm{sys}}$~s~\cite{Elwell2024}) crystals.

The demonstrated fabrication of \thf{} thin film targets with thicknesses 30--100 nm and diameter \SI{50}{\micro\meter}--5~mm and the ensuing observation of the nuclear clock transition therein clearly show a pathway towards a future integrated low-radioactivity solid-state nuclear clock that can be fabricated with existing thin film coating technology. 
In what follows, we describe the \thf{} target fabrication procedure and characterization, the laser spectroscopy apparatus and results, and then discuss the advantage of \thf{} films or \thor{}-doped ThF$_4$ crystals in a solid-state nuclear optical clock.

\medskip
\noindent\textbf{Miniaturized physical vapor deposition of \thf}
\smallskip

\begin{figure*}[h!]
    \centering
    \includegraphics[width=0.9\textwidth]{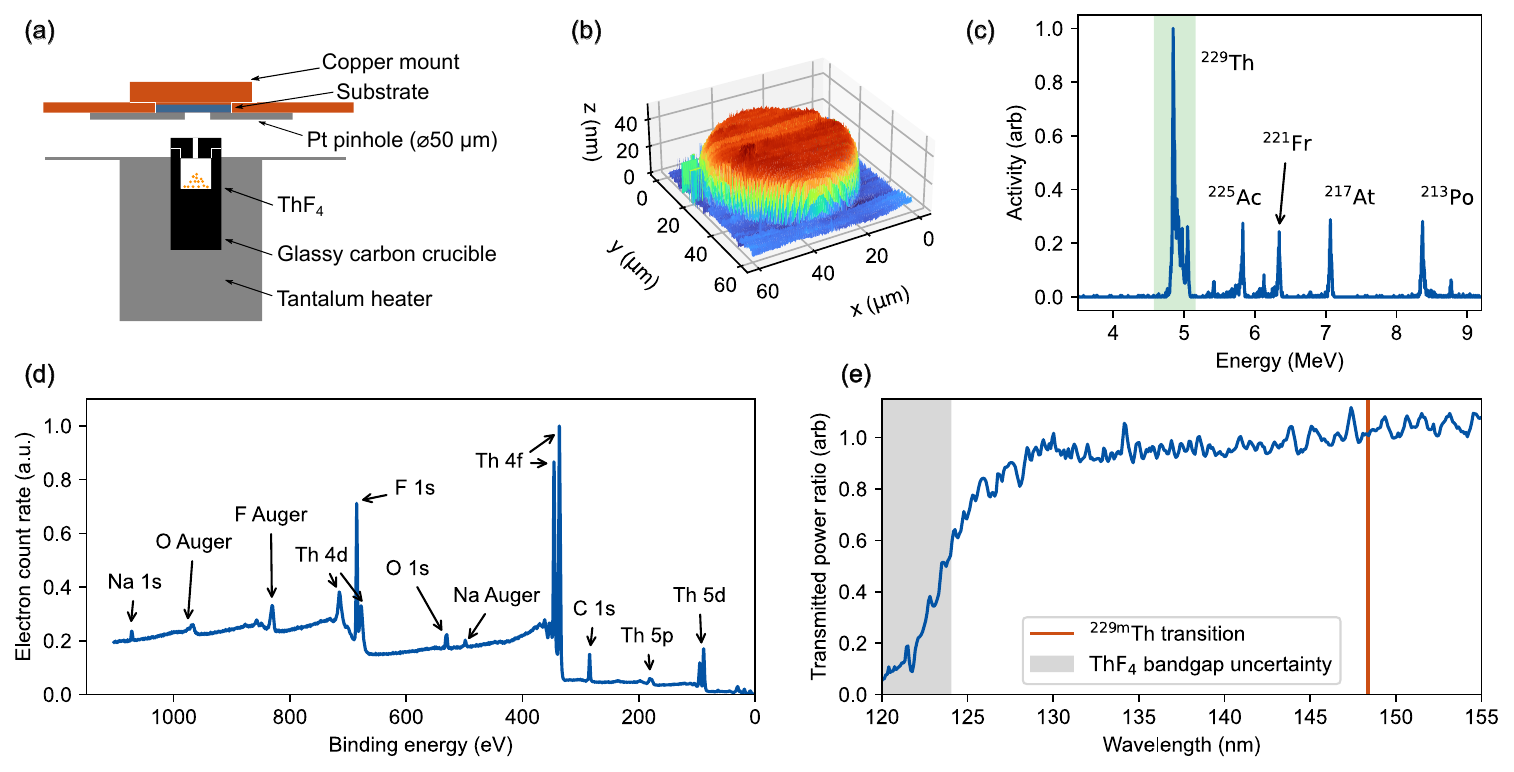}
    \caption{(a)~Schematic of the \thf{} target vapor deposition setup. A small amount ($<$ \SI{10}{\micro\gram}) of \thf{} is loaded into a glassy carbon crucible for radioactive physical vapor deposition. Upon heating, \thf{} vapor effuses from a small ($\sim$ 1 mm diameter) hole on the crucible and deposits on a nearby substrate ($\sim$~3~mm away), forming a $\sim$~5~mm diameter round target. An optional Pt mask is used to generate a small-area target with the desired shape. (b)~Spatial profile of a small \thf{} target measured using AFM. The target is a $\sim$ \SI{50}{\micro\meter} diameter, $\sim$ 30 nm thick disk. (c) Alpha spectrum of the small-area target as shown in (b). Peaks in the green highlighted region (4.59-5.15 MeV) come from \thor{}, corresponding to a \thor{} activity of about 3 Bq. (d) Chemical composition of a similarly fabricated $\sim$ 5 mm diameter \thf{} target measured with XPS. Peaks corresponding to each specific element are annotated~\cite{chastain1992handbook}. This confirms that the thin film's chemical composition is \regthf{}. (e) VUV transmission of a similarly fabricated \regthf{} thin film with $\sim$ \SI{1}{\micro\meter} thickness. The ratio of the transmitted VUV power between the \regthf{} coated and uncoated area on a MgF$_2$ substrate is plotted, confirming the VUV transparency down to the measured \regthf{} bandgap of 10.2(2) eV (shaded gray) in \cite{Gouder2019} and in agreement with the results reported in \cite{Baumeister1973}.}

    \label{fig:vapor_deposition}
\end{figure*}

The vapor deposition setup is shown in Fig.~\ref{fig:vapor_deposition}(a). We start with the \thor{} in dry nitrate form, \thornitr{}, purchased from the National Isotope Development Center. We dissolve the nitrate in ultrapure water for easy handling of microscopic quantities using adjustable volume micropipettes. We directly load the desired quantity of \thornitr{} solution into a glassy carbon crucible and add excess HF to obtain \thf{} precipitate, weighing only a few microgram, that is otherwise impractical to handle. The leftover aqueous solution of HF and HNO$_3$ is then evaporated in a fume hood at elevated temperatures.

After initial material loading, a crucible cap is installed to further decrease the solid angle of the evaporation beam. The crucible is heated via a tantalum heater to above \SI{1000}{\celsius} in 10$^{-6}$ mbar vacuum to vaporize the \thf{} onto a substrate 3 mm away, forming a non-uniform coating of about 5~mm in diameter.
During the coating run, we do not actively control the substrate temperature.
The thickness of this large area coating is measured with a stylus profilometer. The center thickness is typically 30-100 nm, about twice that of the edges, depending on the material amount used in the coating run. The fabricated thin film coatings are likely amorphous (see Methods). 

To fabricate small area targets and minimize the \thor{} consumption, Pt masks with laser-drilled pinholes are used. As Pt has a higher evaporation temperature and is chemically inert, the \thf{} material deposited on the Pt mask can be recycled for future coating runs. The substrate and optional pinholes are held in place with a copper mounting structure (see Fig.~\ref{fig:vapor_deposition}(a)). 

The physical dimensions of one small target, fabricated on a Pt-coated silicon wafer substrate, are measured using an atomic force microscope (AFM) and shown in Fig.~\ref{fig:vapor_deposition}(b).
To demonstrate the capability for customizing target size and shape, we fabricated a $\sim$ \SI{50}{\micro\meter} diameter target using a Pt mask with the same dimensions. With a thickness of $\sim$ 30 nm, the volume of the \thf{} corresponds to roughly $7\times10^{11}$ atoms, or an activity of about 2~Bq. Alpha spectroscopy of the target using a calibrated alpha spectrometer, shown in Fig.~\ref{fig:vapor_deposition}(c), confirms the presence of \thor{} and its daughter isotopes. We extract a ~\thor{} activity of $\sim 3$~Bq from the alpha spectrum, in rough agreement with the estimation from the target volume, by integrating over the region of interest corresponding to the \thor{} alpha peaks (shaded green). 

The chemical composition of the thin film is characterized via x-ray photoelectron spectroscopy (XPS) as shown in Fig.~\ref{fig:vapor_deposition}(d), which confirms the main constituents of the film as thorium and fluorine. Without a detailed analysis of the uncertainties, we extract the atom number percentages from the XPS spectrum to be 19.4\% thorium, 36.6\% fluorine, 34.2\% carbon, 8.4\% oxygen, and 1.4\% sodium. The oxygen and carbon composition presumably originate from hydrocarbon contamination due air exposure. A subsequent XPS measurement on a similar $^{232}$\regthf{} film (not shown) is performed after the removal of the top layer by in-situ argon sputtering. We observe that the peak corresponding to carbon 1s disappears and the oxygen 1s peak decreases significantly, in agreement with this hypothesis. The sodium peak likely comes from trace contamination in the water used during the coating process.

The optical property of the thin film is characterized by measuring the transmitted VUV power ratio between a coated and uncoated section of a single \mgf{} substrate using a D$_2$ lamp and a grating spectrometer. The result is shown in Fig.~\ref{fig:vapor_deposition}(e), where the \thor{} isomer transition wavelength is indicated by the orange line. The spectral transmission agrees qualitatively with the \regthf{} measurement reported in Ref.~\cite{Baumeister1973}, demonstrating high VUV transmission. This data is consistent with the $>10$~eV \regthf{} bandgap predicted in~\cite{Ellis2014} and measured in~\cite{Gouder2019}, as indicated by the gray shaded area. Based on the large bandgap, we expect host quenching effects like internal conversion will not dominate the relaxation of the nuclear clock state, providing the opportunity for radiative fluorescence detection.

\medskip
\noindent\textbf{Nuclear laser spectroscopy}
\smallskip

Using this miniaturized PVD apparatus, two types of large area \thf{} targets (each $\sim$~21~kBq in activity) for the nuclear laser spectroscopy were fabricated on VUV-transparent \sapph{} and \mgf{} substrates.
We performed a coating run for each target with two identical substrates placed side by side, producing two semicircular \thf{} films.
The semicircular \thf{} films are $\sim$~2.5 mm in radius and $\sim$~30 nm in average thickness.
The substrate dimensions are $\sim$~8~mm $\times$ 4~mm $\times~d$~\SI{}{\micro\meter}, where $d=175$~\SI{}{\micro\meter} for \sapph{} and $d=250$~\SI{}{\micro\meter} for the \mgf.
Two identical \thf{} films with the same substrate material are placed facing one another with a $\sim$~1~mm gap between them, forming one target used in the experiment.
This ``sandwich" configuration was chosen to minimize radioactive contamination of the vacuum chamber from recoil daughter nuclei upon \thor{} alpha decay. 

Using the system described in Ref.~\cite{JeetThesis2018, Elwell2024}, VUV radiation was produced via resonance-enhanced four-wave mixing of two pulsed dye lasers in Xe gas. 
The frequency of the first pulsed dye laser, $\omega_u$, was locked to the $5p^{6 ~1}S_0~\rightarrow~5p^5\left(^2P^{\circ}_{3/2} \right) 6p~^2\left[1/2\right]_0$ two-photon transition of Xe at $\sim$ 249.63 nm. 
The frequency of the second pulsed dye laser, $\omega_v$, was scanned to produce VUV radiation in the Xe cell given by the difference mixing relation $\omega = 2\omega_{u} - \omega_v$. 
All three laser beams then impinge off-axis with respect to a MgF$_2$ lens, whose chromatic dispersion is used with downstream pinholes to spatially filter the VUV beam and pass it towards the \thf{} samples. 
The laser system delivers 30 pulses per second to the crystal with a typical VUV pulse energy of $\sim 2$ \SI{}{\micro J}/pulse (see Refs.~\cite{JeetThesis2018, Elwell2024} for details).

\begin{figure*}[h!]
    \centering
    \includegraphics[width=1\textwidth]{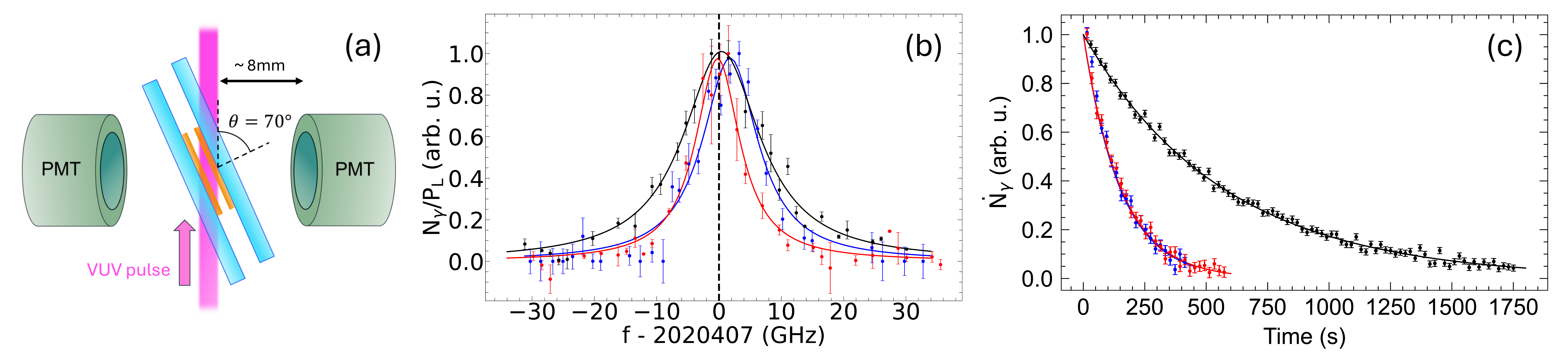}
    \caption{(a)~Schematic of the \thf{} target region. 
    Note that the beam is at a 70$^\circ$ angle with the normal of the two plates to maximize the effective illumination area. 
    The plates are held with a $\sim 1$~mm gap between them.
    (b)~Recorded nuclear fluorescence normalized by laser power versus VUV laser frequency.
    The observed power-normalized fluorescence for the \thf{} film on a \sapph{} (\mgf{}) substrate is shown in blue (red). All error bars represent the standard error of the mean.
    (c) Recorded nuclear fluorescence rate normalized by laser power versus time after the laser is extinguished. 
    The observed radiative decay for \thf{} on a \sapph{}  (\mgf{}) substrate is shown in blue (red). 
    In both (b) an (c) the peak of all curves were normalized to unity and the corresponding result observed in \thor{}:\lisaf{}~\cite{Elwell2024} is shown in black for comparison.  }
    \label{fig:panel_plot}
\end{figure*}

The \thf{} samples were mounted in a vacuum chamber with two VUV-sensitive photo-multiplier tubes (Hamamatsu R6835 PMTs) and a pneumatic shutter system (not shown) to protect the PMTs from direct exposure to the VUV laser, see Fig.~\ref{fig:panel_plot}(a).
The PMTs are operated in a cathode-grounded configuration, and their output waveforms recorded by a 1~Gs/s waveform digitizer (CAEN DT5751) for subsequent post-processing. 
The VUV laser irradiates the samples at approximately 70$^{\circ}$ from the direction normal to the plates to increase the effective illumination volume. 
We terminate the laser beam on a custom VUV energy detector.
The vacuum chamber is maintained with an Ar atmosphere at a pressure of $\sim$10$^{-2}$~mbar to provide high VUV transmission while minimizing degradation of optics due to hydrocarbon deposition~\cite{JeetThesis2018}.

We observe the nuclear transition by monitoring the fluorescence from the \thf{} film following the VUV laser illumination.
Using this system, nuclear spectroscopy of the \thor{} clock transition in the \thf{} film was recorded and is shown in Fig.~\ref{fig:panel_plot}(b). 
We observe a single spectroscopic feature at $2020406.8(4)_\textrm{stat}(30)_\textrm{sys}$~GHz for the \thf{} film on the \mgf{} substrate and at $2020409.1(7)_\textrm{stat}(30)_\textrm{sys}$~GHz for the \thf{} film on the \sapph{} substrate.
For these measurements, the \thf{} was illuminated by the VUV laser for approximately 450~s, the shutters opened in $\sim$~100 ms, and the resulting fluorescence recorded for roughly 450~s. 
Each data point represents an average of the number of photons detected above background in both PMTs and normalized by the total laser power. 
The \sapph{} sample exhibited a lower signal to noise ratio due to an increased radioluminescence background and thus each point was typically repeated around three times.
Shown for reference in Fig~\ref{fig:panel_plot}(b) as black points is the spectra recorded in \thor{}:\lisaf~\cite{Elwell2024}.
The laser bandwidth is estimated to be $\sim 10$ GHz and dominates the linewidth of the observed spectral features.
The smaller spectral linewidth observed in the \thf{} samples as compared to the \thor{}:\lisaf{} sample is attributed to a decrease in laser drift during data collection due to the shorter excitation time here (450~s) as compared to Ref.~\cite{Elwell2024} (1800~s).
The systematic uncertainty of the measurements is dominated by wave meter calibration~\cite{Elwell2024}.

The expected amount of fluorescence is calculated from a detailed analysis of the detection efficiency of our system (see Methods), the known thorium density and thickness of the \thf{} film, and the measured laser parameters, and is found to be considerably larger ($\sim~10^3\times$) than the experimental observation. 
As detailed in Ref.~\cite{Elwell2024}, the hyperfine interaction couples the nuclear degree of freedom to the electronic degree of freedom. 
If there are electronic states with excitations at or below the nuclear energy due to defects in the local crystal structure for example, quenching of the nuclear transition is possible and a reduced fraction $p$ of the \thor{} participate in the narrow nuclear fluorescence.
From the observed signals we find the estimated emitter fraction as $p=2(1) \times 10^{-3}$ for both targets, assuming about $\SI{2}{\micro \joule}$/pulse.

Using the same system, the lifetime of the isomeric transition was measured by collecting the fluorescence as a function of time after laser excitation, see Fig.~\ref{fig:panel_plot}(c), and found to be $150(15)_{\textrm{stat}}(5)_{\textrm{sys}}$~s on \sapph\ and $153(9)_{\textrm{stat}}(7)_{\textrm{sys}}$~s on \mgf.
Here, the systematic error is estimated based on observed drifts in background count rates due to thermal instability of the PMT gain.
These lifetimes are significantly shorter than the value previously observed in \thor{}:CaF$_2$ ($\tau = 641(4)$~s~\cite{Zhang2024}) and \thor{}:\lisaf{} ($\tau =  568(13)_{\textrm{stat}}(20)_{\textrm{sys}}$~s~\cite{Elwell2024}) crystals.

This shortened lifetime is sufficient and perhaps even beneficial for nuclear clock operation. In the SI, we consider multiple mechanisms of lifetime shortening, including the Purcell~\cite{UrbRik1998} and superradiance effects~\cite{Dicke1954,Liao2012}. 
We conclude that the observed shorter isomer lifetime in \thf{} films is likely due to two main factors: a high refractive index of \thf{} and potential quenching effects by the host material. 
While the refractive index required to fully explain the lifetime ($n = 2.2-2.5$) may be higher than our estimate ($n \approx 1.95(30)$),
an additional nuclear decay mechanism can arise 
from a combination of hyperfine mixing of nuclear and electronic states and the hybridization of electronic states due to crystal fields~\cite{Elwell2024}. 
The quenching rate could be further enhanced in thin films like \thf{} due to modified electronic levels~\cite{karpeshin2021} arising from mechanical strain and film-substrate interface effects such as lattice mismatch strains and dipole layer formation at the interface~\cite{Kroemer1975-heterojunction,Brillson2012}.
Other quenching effects from electronic states could arise from the potentially amorphous structure or trace contaminants.

Given these results, it is now clear that the \thor{} nuclear isomeric transition can be driven and observed in \thf{} thin films and potentially even bulk crystals.
This opens many exciting possibilities as \regthf{} film growth is well-developed for e.g. optical coatings, meaning that \thor{} can be directly incorporated in optical components, such as mirrors, lenses, and waveguides.
Thus, a \thf{} nuclear clock is a promising direction for development and therefore in what follows we analyze the expected performance of a \thf-based nuclear clock.

\medskip
\noindent\textbf{Projected \thf{} clock performance}
\smallskip

\begin{figure*}[h!]
    \centering
    \includegraphics[width = 0.8\textwidth]{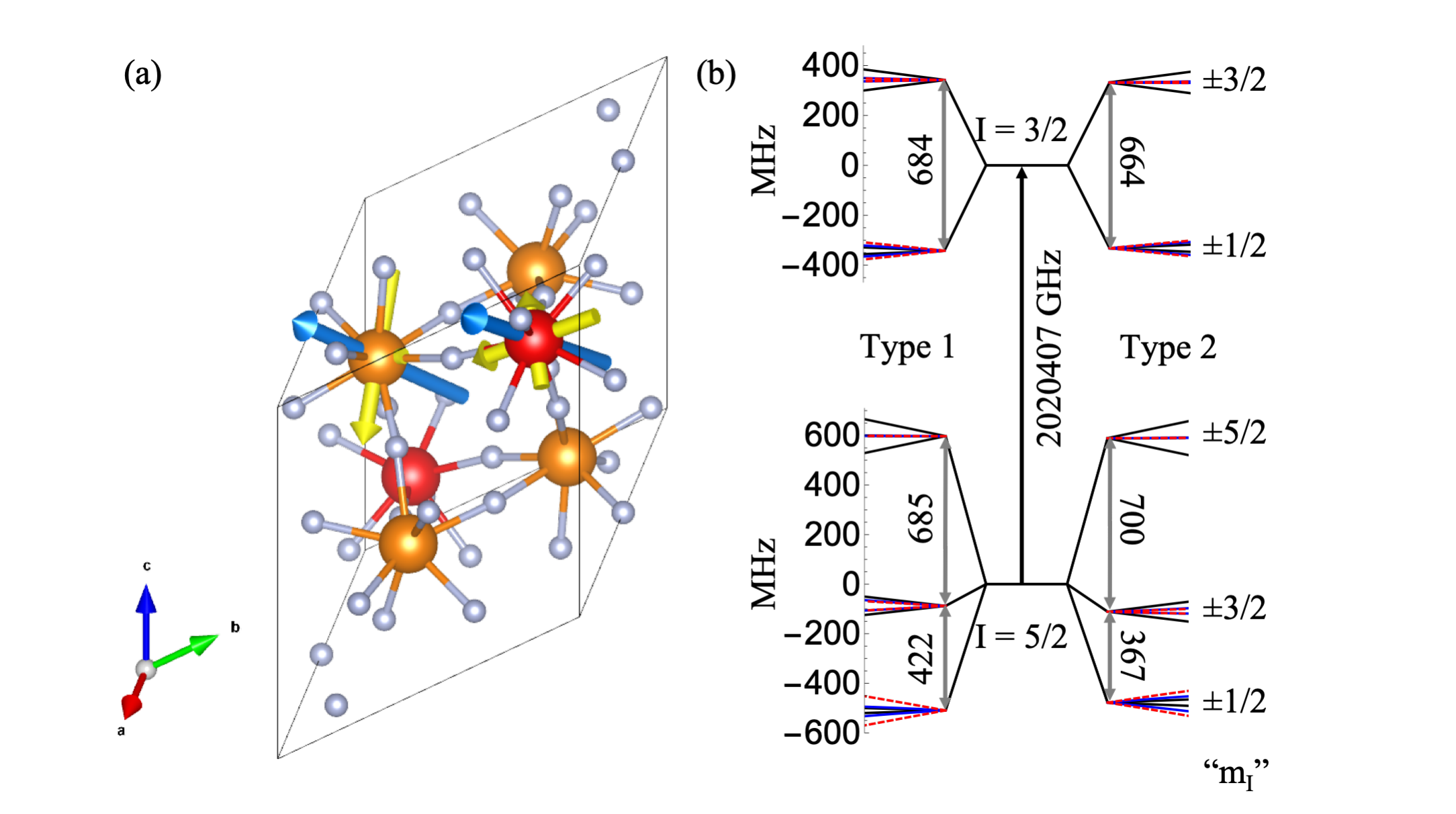}
    \caption{(a) Unit cell of \ce{ThF4} showing electric field gradient (EFG) eigenvectors. Thorium atoms are shown as red and orange spheres, the two colors denoting the two inequivalent crystallographic positions, and fluorine atoms are shown in grey. The light blue vectors show the largest diagonal EFG components, $V_{zz}$, and the yellow vectors show the minor components, $V_{xx}$ and $V_{yy}$. The lengths of the vectors are proportional to the EFG magnitudes along the respective directions. (b) Numerically calculated EFG and Zeeman splittings in the two inequivalent thorium sites. The Zeeman shift from 0~T to 10~T is plotted for a magnetic field along the $\hat{x}$ (blue), $\hat{y}$ (red, dashed), and $\hat{z}$ (black) direction.}
    \label{fig:xtal_splitting}
\end{figure*}

While the crystallinity of the fabricated \thf{} thin films needs to be characterized (see Methods) and possibly improved~\cite{beeks2024} in future experiments, it is well known that \regthf{} single crystals~\cite{pastor1974preparation} can be grown. Therefore, we perform density functional theory (DFT) calculations (see Methods) using the monoclinic crystal structure of ThF$_4$, space group $C2/c$ (\#15)~\cite{RN566}, (see Fig.~\ref{fig:xtal_splitting}(a)) in which the thorium atoms are nominally in the \thor{}$^{4+}$ charge state. 
The main effects of the crystal on the \thor{}$^{4+}$ nucleus is the coupling of the \thor{} nuclear electric quadrupole (EQ) moment to the crystal electric field gradients (EFG) $\{V_{xx},V_{yy},V_{zz}\}$ and the coupling of the \thor{} nuclear magnetic moment to the magnetic field created by the other atoms in the crystal~\cite{Rellergert2010}. 
\thf{} is an ionic crystal with no unpaired electrons and therefore the magnetic field experienced by a \thor{} atom comes from the other nuclear magnetic moments in the crystal. 
Thus, the nuclear energy levels are described by the Hamiltonian:
\begin{equation}
    \hat{H}= -\mu_\alpha \vec{I}\cdot\vec{B} + \frac{eQ_\alpha V_{zz}}{4I(2I-1)}\left[3\hat{I}_z^2-\hat{I}^2 +\eta(\hat{I}_x^2-\hat{I}_y^2)\right], \nonumber
\end{equation}
where $\hat{I}$ is the total nuclear spin operator, $\hat{ I}_{x,y,z}$ are the component of the nuclear spin operators, $\eta = |V_{xx}-V_{yy}|/V_{zz}$ is the EFG asymmetry parameter with the choice $|V_{zz}| > |V_{xx}| > |V_{yy}|$, and $\alpha = \{g,e\}$ denotes the ground and excited nuclear states, respectively, which is parameterized by $Q_g = 3.149(4)$~eb~\cite{Bemis1988}, $\mu_g = 0.360(7)\mu_N$~\cite{Gerstenkorn1974, Campbell2011}, $Q_e = 1.77(2)$~eb~\cite{Thielking2018, Yamaguchi2024}, and $\mu_e = -0.378(8)\mu_N$~\cite{Thielking2018, Yamaguchi2024}, where $\mu_N$ is the nuclear magneton.

In contrast to the \thor{}-doped high bandgap crystals previously considered~\cite{Jackson2009,Rellergert2010b,Pimon2022}, \regthf{} is expected to host \thor{} in two non-equivalent sites~\cite{RN566}.
The first site (Type 1), shown as orange spheres in Fig.~\ref{fig:xtal_splitting}(a), experiences a crystal EFG of $V_{zz}=310.5$~V/\AA$^2, \eta=0.437$ and is twice as populated as the second site (Type 2), shown as red spheres in Fig.~\ref{fig:xtal_splitting}(a), which experiences a crystal EFG of $(V_{zz}=308.9$~V/\AA$^2, \eta=0.194)$.

Due to the differing electric field gradients, these two sites have non-degenerate energy levels (see Fig.~\ref{fig:xtal_splitting}(b)) and therefore distinct laser spectra.
Having control of two separate populations of \thor{} nuclei in this crystal enables a number of new experiments including study of spin exchange interactions between thorium nuclei, and could be used to improve the performance of a nuclear clock as these populations can be independently probed, thereby reducing the Dick effect~\cite{Ludlow2015}. 

A major source of inhomogeneous broadening of the nuclear clock transitions is magnetic dipole-dipole interaction between the \thor{} nuclei as well as with the nearby $^{19}$F nuclei.
By using the positions found for the \thor{} and $^{19}$F atoms from the DFT calculation, the magnetic moments of the nuclei, and assuming random orientation of the nuclear spins, we estimate that the magnetic field seen by a thorium nucleus fluctuates with a standard deviation of approximately 5~G. 
Thus, the Zeeman interaction broadens the levels by an amount of order 1-10~kHz, while the transition linewidth depends on the relative Zeeman shift of the states involved in the clock transition. 
At first glance, because the magnetic moment of the \thor{} isomeric state is nearly equal but opposite to that of the nuclear ground state, a clock based on, for example, either of the $\ket{I = 5/2, m_I = \pm1/2} \leftrightarrow \ket{3/2,\mp1/2}$ transitions would exhibit a magnetic field sensitivity of only $0.009(5)\mu_N  = 7(4)$~Hz/G, leading to a broadening of $\lesssim~100$~Hz. 
However, because the EFG orients the nucleus, the Zeeman sensitivity depends strongly on the direction of the magnetic field relative to the EFG axes, as shown in Fig.~\ref{fig:xtal_splitting}(b) where the Zeeman shift from 0 to 10 T is plotted for a magnetic field along the $\hat{x}$ (blue), $\hat{y}$ (red, dashed), and $\hat{z}$ (black) direction. 
In this figure the energy levels are labelled by the $m_I$ quantum number of the primary component of the eigenstate. 
The alignment of the quadrupole moment, and therefore the nuclear spin and magnetic moment to the EFG $\hat{z}$ axis, results in an increased (decreased) sensitivity of stretched states to magnetic fields along the $\hat{z}$ axis ($\hat{x}$ and $\hat{y}$ axes).
This significantly alters the Zeeman sensitivity of the available nuclear transitions.
By numerically diagonalizing the Hamiltonian for a sample of magnetic fields (see Methods) the Zeeman limited linewidths of the various transitions are estimated to range between approximately 1~kHz -- 7~kHz, with the least broadening of $\approx 1.1$~kHz realized by the $\ket{5/2,\pm1/2}\leftrightarrow\ket{3/2,\mp1/2}$ transition for Type 2 sites. 
This increased Zeeman sensitivity as compared to a free \thor{} nucleus highlights the role the EFG plays in determining clock performance and suggests that a crystal structure with vanishing EFG, as would occur if \thor{} was positioned in a site of cubic symmetry, could lead to improved performance. 

Because the quantization axis is set by the crystal EFG, selecting a nuclear transition with improved magnetic field performance requires light of specific polarization propagating in a direction set by the crystal EFG. 
This is straightforward in a single crystal of \thf, but may be more difficult to achieve in \thor{}:\lisaf{}~\cite{Rellergert2010a, Elwell2024} and \thor{}:CaF$_2$~\cite{Tiedau2024} since the symmetry axis of the electric field gradient depends on the geometry of the interstitial $^{19}$F atoms required for charge compensation. 

With these calculations, the performance of a \thf{} thin film nuclear clock can be estimated. 
Assuming a 100~nm thick film, a probe laser linewidth significantly smaller than the inhomogeneous Zeeman-limited transition linewidth, and probe laser power of \SI{1}{\micro W}, the performance of a clock based on the $\ket{5/2,\pm1/2}\leftrightarrow\ket{3/2,\mp1/2}$ transition is estimated to have a fractional instability of $5\times10^{-17}$ at \SI{1}{s} for both Type 1 and Type 2 sites, see Methods.
Estimating the inaccuracy is not yet possible for these systems and will be the subject of future work, however, strain and temperature sensitivity~\cite{Rellergert2010} can be expected as important systematics. 

\medskip
\noindent\textbf{Discussion and outlook}
\smallskip

Besides precision measurement applications, the vapor deposited \thf{} thin films with controllable thickness also provides a new platform for studying Purcell effects. We expect when the film thickness is comparable to or smaller than the excitation wavelength, the film thickness and substrate refractive index can be exploited to control the emitter lifetime and emission direction. These effects can potentially be used for engineering new quantum optics devices or improving the nuclear clock performance.

While superradiant effects are not observed in this work due to the inhomogeneous broadening and low participation factor, the emitter density in \thf{} $(\lambda/n_{\textrm{ThF}_4})^3 \rho_\textrm{Th} > 10^6$ is more than 3 orders of magnitude higher than that achieved in \thor{}-doped crystals.
Using \thf{} waveguides or resonantors for increased optical density, a new regime for quantum optics studies involving nuclear superradiance~\cite{rohlsberger2010} and coherent nuclear forward scattering~\cite{Liao2012} appears accessible in \thf.

With this demonstration of nuclear spectroscopy in \thf{} thin films, we also rekindle the interest of the community in \regthf{} crystals. 
Doping \thor{} in \regthf{} does not create defects or alter the crystal structure.
This uniquely enables control over the crystal EFG quantization axis for polarization-selective excitation of the clock states. 
Also, in \thor{}-doped thin film or crystalline \regthf{} samples, one can tune the emitter density for precision measurement and quantum optics studies. 
Moreover, \regthf{} multilayer coatings have been used regularly in industrial applications, promising a clear route to scalable production of solid-state nuclear clocks based on integrated \regthf{} photonic structures that can be incorporated with VUV lasers and detection systems.

As the observed participation fraction of the nuclear transition in optical decay is low ($\sim 10^{-3}$), it is necessary to continue investigating the underlying quenching mechanism and methods for improvement.
Narrow, state-resolved nuclear transitions~\cite{Zhang2024} in the fabricated thin films have not yet been observed, due to the low participation fraction and potentially also inhomogeneous line-broadening effects caused by amorphous characteristics of the sample.
Annealing and fluorination processes may significantly increase the participation fraction and microcrystalline structure, as demonstrated both in M\"ossbauer spectroscopy measurements~\cite{dornow1979} and \thor{}-doped CaF$_2$ crystals~\cite{beeks2024}. 
 Single crystal $^{232}$\regthf{} substrates~\cite{pastor1974preparation} can potentially be used for epitaxial \thf{} film growth.

Looking forward, the quenching of the isomer excited state in a thin film can also be exploited for constructing a nuclear clock based on internal conversion electrons~\cite{vonderwense2020concept} with comparable performance to clocks based on fluorescence photons. 
The \regthf{} thin film can be converted to other chemical forms, such as ThO$_2$ with a cubic lattice structure (see Methods), for investigation of the quenching effect and for building a conversion-electron-based clock. The electron escape probability from a ThO$_2$ thin film may limit the detection efficiency and needs to be studied.

In summary, we address the challenge in the fabrication of solid-state nuclear clocks by demonstrating VUV nuclear spectroscopy of the \thor{} clock transition in \thf{} thin films grown by miniaturized radioactive PVD. 
Based on detailed DFT calculations, we predict the performance of a \thf{} solid-state nuclear clock. 
Solving important obstacles of material availability and radioactivity limits, our work lays the foundation for future scalable production and widespread use of low-cost integrated nuclear clocks, towards the promise of a simple, portable frequency reference. 

\clearpage

\clearpage

\putbib[ref]
\end{bibunit}

\bigskip
\textbf{Data availability}

The data that support the findings of this study are available from the corresponding author upon appropriate request.

\medskip
\textbf{Acknowledgement}

We acknowledge funding support from the Army Research Office (W911NF2010182); Air Force Office of Scientific Research (FA9550-19-1-0148); National Science Foundation QLCI OMA-2016244; National Science Foundation PHY-2317149; and National Institute of Standards and Technology. J.S.H. acknowledges support from the National Research Council Postdoctoral Fellowship. L.v.d.W. acknowledges funding from the Humboldt Foundation via a Feodor Lynen fellowship. We thank the National Isotope Development Center of DoE and Oak
Ridge National Laboratory for providing the Th-229 used in this work.

This work was also supported in part by NSF awards PHYS-2013011, PHYS-2412982, and PHY-2207546, and ARO award W911NF-11-1-0369.
E.R.H acknowledges institutional support by the NSF QLCI Award OMA-2016245.
This work used Bridges-2 at Pittsburgh Supercomputing Center through allocation PHY230110 from the Advanced Cyberinfrastructure Coordination Ecosystem: Services \& Support (ACCESS) program, which is supported by National Science Foundation grants \#2138259, \#2138286, \#2138307, \#2137603, and \#2138296.

We thank Christoph Düllmann, Dennis Renisch for helpful discussions, Junyu Lin, Dileep Reddy for proofreading, Andrew Cavanagh, Curtis Beimborn, Kim Hagen, Viki Martinez, and the CU Radiation Safety Office for technical assistance. We acknowledge use of the Materials Research X-Ray Diffraction Facility at the University of Colorado Boulder (RRID: SCR\_019304), with instrumentation supported by NSF MRSEC Grant DMR-1420736. 

L.v.d.W.’s present address is Johannes Gutenberg-Universität Mainz, Institut für Physik, Staudingerweg 7, 55128 Mainz, Germany. H.U.F.'s present address is Am Heckenacker 40, D-8547 Wörth, Germany.

\medskip
\textbf{Author contributions}

C.Z., L.v.d.W, J.F.D., J.S.H., T.O., H.U.F, J.Y. produced and characterized the sample. R.E., J.E.S.T., E.R.H. performed the laser spectroscopy measurement. H.W.T.M., A.N.A., H.B.T.T., A.D. performed theory calculations. All authors wrote the manuscript.

\medskip
\textbf{Competing interests}

The authors declare no competing interests.

\medskip
\textbf{Additional information}

Correspondence and requests should be addressed to J.Y. (ye@jila.colorado.edu) and E.R.H. (eric.hudson@ucla.edu).

\newpage

\begin{bibunit}
\newbibstartnumber{56}

\section{Methods}

\subsection{Verifying the VUV Indices of Refraction}
Data on indices of refraction in the VUV are sparse, and so it is useful to verify the values obtained from literature. To do this, samples of the various materials used in our experiment were placed at normal incidence in our VUV beamline in order to measure their transmission at $\sim$148.38 nm. 

Literature values for the refractive indices of \mgf~\cite{Li1980} are 1.4828 (O-ray) and 1.4977 (E-ray), while the value for \sapph\ a value of 2.32(2) is obtained by averaging various analyses found in~\cite{French1998}. A thin sample of \mgf\ ($\SI{250}{\micro\meter}$) was found to have $91(4)\%$ transmission, while a thin sample of \sapph\ ($\SI{175}{\micro\meter}$) $61(9)\%$. Assuming no additional absorption and normal Fresnel reflection losses, this would be consistent with values of refractive indices of $n=1.55(17)$ for \mgf\ and $n=2.77(41)$ for \sapph. Thus our transmission measurements are in agreement with the literature values of the indices of refraction assuming no additional absorption in the thin samples.

Obtaining literature values for the VUV refractive index of ThF$_4$ is more challenging as far less data is available. 
Extrapolating the measurements taken in~\cite{Baumeister1973} to 148.38 nm leads to a prediction of $n = 1.82$ for ThF$_4$. 
As the index of refraction typically rises dramatically in the VUV as the band edge of the material is approached, this value is likely a lower bound.
Therefore, we also performed a rudimentary measurement of $n_{\textrm{ThF}_4}$ by comparing the amount of 148.38 nm light transmitted through a 5 mm thick sample of \mgf{} that was uncoated and coated with a thin \thf{} film. 
By assuming no absorption in the thin \thf\ film and normal Fresnel reflections at the interfaces, a value of $n_{\textrm{ThF}_4} = 1.95(30)$ was obtained.

\subsection{Calculation of the System Detection Efficiency}
The system detection efficiency is comprised of two main components: the efficiency with which a photon emitted from the decay of the isomer is collected by our detectors and the quantum efficiency (QE) of the detectors themselves. 
The QE of our PMT detectors (Hamamatsu model R6835) was provided by the manufacturer's calibrations ($\approx 12 \%$ at 148nm). 

The collection efficiency is calculated from custom ray-tracing software. 
This program accounts for refraction and multiple reflections in the \ce{ThF4}-substrate system to provide an accurate distribution of rays emerging from the sample. 
Given that the \thf{} samples are inhomogeneous in thickness, we make the approximation that interference effects are negligible. 
The efficiency with which rays emerge from the \sapph{} (\mgf) substrate towards the respective PMT is $\approx 7.6\%$ ($\approx 8.6\%$).
The calculated ray distributions from the sample into vacuum are passed to a commercial ray-tracing software to determine the number of rays that ultimately reach the detectors~\cite{zemax2012}, which was $\approx 72 \%$ for both substrates. 
As such, the total system detection efficiency was $\approx 0.7 \%$ for both substrates.

\subsection{Signal Detection and Backgrounds}
The output waveform pulses from the R6835 PMT are amplified through a Stanford Research Systems SR445A preamplifier before being routed to the waveform digitizer. The digitizer is set to trigger at $\approx 25\%$ pulse amplitude and has an $\sim$~8 ns dead time between waveform acquisitions. In both \thf{} targets on \sapph{} and \mgf{}, the total number of detected signal photons from the nuclear isomer decay was of the order $\sim$ 2$\times 10^4$ (varying with laser power), while the backgrounds were $\sim$~9000 cps (1300 cps) for \sapph{} (\mgf{}). cps, counts per second.

\subsection{Additional $^{229}$T\MakeLowercase{h}F$_4$ thin film properties}
The \thor{} activity of both \thf{} targets used in the nuclear laser spectroscopy was measured to be $\sim 21$~kBq, corresponding to $\sim 7.5 \times 10^{15}$ \thor{} nuclei. We use the known activity of \thor{} ($1.68\times10^{12}$ Bq/mol), the density of \thf{} (6.3 g/cm$^{3}$), and the diameter of the target ($\sim$ 5mm) to roughly estimate the thickness of the films. The thickness obtained is approximately 30 nm. 
This is in agreement with profilometer and AFM measurement results for different targets, scaled by their radioactivity.

Thin film targets created by vapor deposition tend to be amorphous. We performed preliminary studies of the thin film structure using grazing-incidence wide-angle scattering X-ray diffraction. The correlation length, which is an estimation of the average grain size, is extracted using the Scherrer equation~\cite{steele_giwaxs_2023} to be about \SI{30}{\angstrom} for a similarly produced $^{232}$\regthf{} sample. In future studies, X-ray diffraction measurements in combination with VUV nuclear spectroscopy will help characterize the \thf{} nuclear transition sensitivity to the sample crystallinity.

\subsection{Density Functional Theory calculations}

Density functional theory (DFT) calculations were performed using the PBE (Perdew, Burke, Ernzerhof) functional and the projector augmented wave (PAW) method as implemented in the Vienna Ab initio Simulation Package (VASP), version 6.4.2~\cite{RN12,RN13,RN14}.
Calculations were done on the primitive cell of \ce{ThF4}, which contains six formula units.
All calculations used a 500 eV plane wave cut-off and a $\Gamma$-centered $k$-mesh with a $k$-point spacing of 0.03 \AA{}$-1$ (4-4-5 for the primitive unit cell).
The structure (atomic positions and lattice parameters) of \ce{ThF4} was optimized until the force components on all atoms were less than 0.02 eV/\AA{}.
Electric field gradients were computed using the method of Petrelli \textit{et al.} implemented in VASP~\cite{RN538}.
Convergence tests with respect to plane wave cut-off and SCF (self-consistent field) convergence criterion were performed up to 800 eV and $10^{-8}$~eV respectively.
The tests showed that settings of 500 eV and $10^{-5}$~eV, used in all our calculations, were sufficient to converge the electric field gradients.

A single-point calculation with the HSE06 hybrid functional~\cite{RN589} was used to confirm the accuracy of the computed electric field gradients.
The parameters for the first site are $V_{zz}=309.3$~V/\AA$^2, \eta=0.428$ and for the second site
$V_{zz}=308.1$~V/\AA$^2, \eta=0.194$, matching the PBE results very well.
This calculation used the same computational settings as above except that a 4-4-4 $k$-mesh was used with a downsampling factor of 2 for the Fock operator.
The band gap computed with HSE06 was 8.85 eV, consistent with previous calculations~\cite{Ellis2014}.

\subsection{Nuclear transition quenching in thin films}

Our measured $\sim$ \SI{160}{s} lifetime of the $^\mathrm{229m}$Th isomeric excited state in \thf{} films is significantly shorter than $\sim 600$~s $^\mathrm{229m}$Th lifetime measured in doped crystals~\cite{Tiedau2024,Elwell2024}. 
The nuclear lifetime can be affected by multiple factors and quenching mechanisms.

The bare $^\mathrm{229m}$Th isomer predominantly decays to its ground state by magnetic-dipole (M1) radiation. 
The corresponding vacuum isomer lifetime is $\tau_{is} \sim 1900-2500$~s~\cite{Tiedau2024,Yamaguchi2024,Elwell2024,Kraemer2023}. 
The \thor{} nucleus in ThF$_4$, however, is embedded into the electronic cloud of Th$^{4+}$ ion and the ion itself into the crystal lattice. 
The M1 decay rate can be modified by the host environment and new, competing, decay channels may open up. 

We start by reminding the reader that if a quantum emitter is embedded into a bulk non-magnetic dielectric with refractive index $n$, its  M1 decay rate   is increased by a factor of $n^3$ compared to its vacuum value $A_{M1}^\mathrm{vac}$~\cite{Nienhuis1976a,Tkalya2000-M1inMedium}. 
The M1 radiative lifetime becomes shorter in a dielectric, due to photon mode field renormalization and the change in the photon density of states. 
This is an accepted explanation for shortened isomer lifetime in doped crystals~\cite{Tiedau2024,Elwell2024} as compared to its vacuum (bare nucleus) value~\cite{Yamaguchi2024}.
In our experiment, however, \thf{} comes as a film deposited on a substrate.  
The the substrate-film and vacuum-film interfaces modify the allowed photon modes and, therefore, the coupling and the photonic density of states. 
This influences both the emission rate and angular distribution~\cite{UrbRik1998,Kerker1978}.  Following Ref.~\cite{UrbRik1998}, we quantized electromagnetic fields in our experimental geometry. 
We find that  the \thor{}  M1 rate can be parameterized as $A_{M1}(\mathbf{r}) = g(\mathbf{r} ) A_{M1}^\mathrm{vac}$, where the position-dependent M1 Purcell factor  $g(\mathbf{r} )$ scales the vacuum $A_{M1}^\mathrm{vac}$ rate. 
For a bulk dielectric with refractive index $n$, $g(\mathbf{r} ) = n^3$. 
The M1 Purcell factor for non-magnetic media is continuous across interfaces. 
Therefore, $g(\mathbf{r})$ in a thin film ($d \ll \lambda/n_{\textrm{ThF}_4}$) must smoothly interpolate between the bulk substrate value $n_\mathrm{sub}^3$ and the vacuum value of 1. 
Thus, the M1 decay rate depends on how deep a \thor{} nucleus is in the film, and an experiment would not observe a single exponential decay. 
In addition, the rate would depend on the substrate refractive index. 
In the opposite limit of thick films, we find that  $g(\mathbf{r} ) \approx n_{\textrm{ThF}_4}^3$ and is largely position and substrate-independent, resulting in a single exponential decay; this is consistent with data in Fig.~\ref{fig:panel_plot}(c). 
Moreover, our measured decay rates for \thf{} are indistinguishable for decay on the \mgf{} and \sapph{} substrates, while these two substrates have substantially different refractive indexes. 
We conclude that the Purcell effect due to the substrates does not play a leading order role in determining the isomer lifetime in our experiments.

The emitter lifetime can be also shortened by collective coherent effects like superradiance~\cite{Dicke1954,Liao2012}. 
These appear unlikely, since the thorium number density per wavelength cubed, $(\lambda/n_{\textrm{ThF}_4})^3 \rho_\textrm{Th} > 10^6$, is reduced by the inhomogeneous broadening~\cite{Rellergert2010a}, $\Gamma_{nat}/\Gamma_{inh} < (1/150)~\textrm{s}/(2\pi\times1~\textrm{kHz}) = 10^{-6}$ and the participation factor, $p= 10^{-3}$. 

Other explanations for the shorter isomer lifetime observed here include a large index of refraction for ThF$_4$ and quenching by the host material. 

Explaining the observed lifetime as a consequence of the \thf{} refractive index requires $n_{\textrm{ThF}_4} = \sqrt[3]{\tau_{is}/\tau} = 2.2-2.5$, given the vacuum isomer lifetime $\tau_{is} \approx 1800-2500$~s~\cite{Tiedau2024,Elwell2024}. 
To our knowledge, the shortest wavelength measurement of $n_{\textrm{ThF}_4}$ is down to 157~nm~\cite{Baumeister1973}. 
Linearly extrapolating this result to 148~nm predicts $n_{\textrm{ThF}_4} = 1.82$; this estimate is likely a lower bound as the index of refraction of a material typically rises dramatically in the VUV.
To provide an additional estimate of $n_{\textrm{ThF}_4}$ at 148~nm, we measured and compared the transmission of our laser through a \thf-coated \mgf{} substrate to that of an uncoated \mgf{} substrate.
Assuming the transmission is determined solely by the Fresnel reflection at the interfaces, this measurement suggests $n_{\textrm{ThF}_4} = 1.95(30)$.
Though this value is close to the $n_{\textrm{ThF}_4}$ inferred from the isomeric lifetime, it may not fully explain the observed lifetime.  

Ref.~\cite{Elwell2024} has pointed out a novel electric-dipole (E1) channel for a nuclear decay in a solid-state environment. 
This decay mechanism is enabled by hybridization of electronic states by the crystal fields. 
The hyperfine interaction couples the nuclear degree of freedom to these mixed-parity electronic states, allowing for a competing E1 decay channel for the isomeric state. 
Qualitatively, in thin \thf{} films, the mixing of opposite-parity electronic states can be further enhanced by electric fields generated due to internal mechanical strains. 
Interface between \thf{} and the substrate can also generate large mixing electric fields: (i) lattice mismatch at the interface can cause strains~\cite{Brillson2012}, and (ii) when two dissimilar insulators are brought into contact, their Fermi levels must align causing ``band bending'' and space charge and dipole layer formation~\cite{Kroemer1975-heterojunction,Brillson2012}.   

In addition to the enumerated quenching mechanisms, the nuclear transition lifetime may be affected by the bulk and surface impurities and stray surface charges.

While the observed shorter isomer lifetime is not yet fully understood, it also offers a rich landscape for exploration at the interface of nuclear and solid-state physics and quantum optics, potentially leading to new insights in these fields.
This will be the subject of future studies.

\subsection{Zeeman Sensitivities of the Nuclear States}

The nuclear energy levels are described by the Hamiltonian:
\begin{equation}
    \hat{H}= -\mu_\alpha \vec{I}\cdot\vec{B} + \frac{eQ_\alpha V_{zz}}{4I(2I-1)}\left[3\hat{I}_z^2-\hat{I}^2 +\eta(\hat{I}_x^2-\hat{I}_y^2)\right], \nonumber
\end{equation}
where $\hat{I}$ is the total nuclear spin operator, $\hat{ I}_{x,y,z}$ are the component of the nuclear spin operators, $\eta = |V_{xx}-V_{yy}|/|V_{zz}|$ is the EFG asymmetry parameter, and $\alpha = \{g,e\}$ denotes the ground and excited nuclear states, respectively, which is parameterized by $Q_g = 3.149(3)$~eb~\cite{Bemis1988}, $\mu_g = 0.360(7)\mu_N$~\cite{Gerstenkorn1974, Campbell2011}, $Q_e = 1.77(2)$~eb~\cite{Yamaguchi2024}, and $\mu_e = -0.378(8)\mu_N$~\cite{Yamaguchi2024}, where $\mu_N$ is the nuclear magneton.
\begin{table}[h!]
    \centering
    \begin{tabular}{c|c|c|c}
         $\ket{3/2,``m_I"}$ & $\ket{5/2,``m_I"}$  & $\Gamma/(2\pi)$~kHz &   $\Gamma/(2\pi)$~kHz \\
          &  & (Type 1) &   (Type 2) \\
         \hline 
         $\pm 3/2$ &  $\pm 5/2$ & 6.5 & 6.5\\
         $\pm 3/2$ &  $\pm 3/2$ & 5.4 & 5.1 \\
         $\pm 3/2$ &  $\pm 1/2$ & 6.1 & 6.1\\
         $\pm 1/2$ &  $\pm 3/2$ & 5.7 & 5.3\\
         $\pm 1/2$ &  $\pm 1/2$ & 6.8 & 6.9\\
         $\pm 1/2$ &  $\mp 1/2$ & 1.3 & 1.1\\
    \end{tabular}
    \caption{Estimated linewidths of the allowed M1 transitions.}
    \label{tab:my_label}
\end{table}

\regthf{} hosts \thor{} in two non-equivalent sites. 
The first site (Type 1), shown as orange spheres in Fig.~\ref{fig:xtal_splitting}(a), experiences a crystal EFG of $(V_{zz}=310.5$~V/\AA$^2, \eta=0.437)$ and is twice as populated as the second site (Type 2), shown as red spheres in Fig.~\ref{fig:xtal_splitting}(a), which experiences a crystal EFG of $(V_{zz}=308.9$~V/\AA$^2, \eta=0.194)$.
Due to the differing electric field gradients, these two sites have non-degenerate energy levels (see Fig.~\ref{fig:xtal_splitting}(b)) and therefore distinct laser spectra as shown in Fig.~\ref{fig:xtal_splitting}(b).

To calculate the expected linewidth of the various \thor{} nuclear transitions, we perform a Monte Carlo simulation as follows. 
First, it is assumed that the nuclear spin of each $^{19}$F and \thor{} atom is randomly oriented and the magnetic field due to these magnetic dipoles created at the central thorium nucleus is calculated.
For this calculation, the structure shown in Fig.~\ref{fig:xtal_splitting}(a) is extended to include $\sim$~500 atoms to ensure convergence. 
This process is repeated 10,000 times to create a sample of possible magnetic fields at the central thorium nucleus. 
Next, the Hamiltonian above is numerically diagonalized for both $I = 5/2$ and $I = 3/2$ states at each sample of the magnetic field and the shift in transition frequency is found. 
The average and the standard deviation of the sample of transition frequency shifts is found.
The average shift is approximately zero, as expected since the EFG defines the quantization axis, while the standard deviation, which should be a reasonable estimate of the transition linewidth, is shown in Table~\ref{tab:my_label}.

\subsection{Conversion from fluoride to oxide}
As mentioned in the main text, a \thox{}  thin film is desirable for its cubic symmetry leading to vanishing electric quadrupole splitting as well as for constructing a conversion-electron-based nuclear clock due to its low bandgap~\cite{vonderwense2020concept}. 
We have developed a procedure for reliably converting a \thf{} thin film to \thox{} using pyrohydrolysis. 

For the conversion, the \thf{} target is placed on a hot plate (Corning, PC-420D) set to a temperature of 550$^\circ$C. 
Argon gas is saturated with water by being sent through a water-filled bubbler and is flowed through a glass tube positioned directly over the target. 
In addition, a 300~W halogen lamp (Omnilux ELH, 120~V) is placed a few centimeters from the target to further increase the surface temperature. The process is terminated after about one hour, although significantly shorter reaction times might be sufficient. 
Under these conditions the fluoride layer converts to oxide following the reaction
$$
\text{ThF}_4 + 2 \text{H}_2\text{O} \xrightarrow[\hspace{1cm}]{} \text{ThO}_2 + 4 \text{HF} .
$$
It was a concern that the surface layer might disintegrate during the process of pyrohydrolysis and that the oxide layer might not stay attached to the substrate. 
Fortunately, no such problems occurred. However, shrinking of the layer thickness, as expected for stoichiometric reasons, did occur and was confirmed via AFM measurements. 
The chemical composition of the target was also confirmed as \thox{} through XPS. 
Notably, pyrohydrolysis is a very aggressive process, potentially also leading to the oxidation of the substrate. 
The laser excitation of a \thox{} thin film, possibly observed by monitoring conversion electrons~\cite{vonderwense2020concept}, has yet to be demonstrated and will be the subject of future studies.

\subsection{Determination of Clock Stability}
We estimate the clock stability by~\cite{MartinBoydThesis2007}
\begin{equation*}
    \sigma = \frac{1}{2 \pi Q S}\sqrt{\frac{T_e + T_c}{\tau}},
\end{equation*} where $Q=f_0/\Delta f$ is the transition quality factor, $S$ in the signal-to-noise ratio (SNR), $T_e$ is the excitation time, $T_c$ is the fluorescence collection time, and $\tau$ is the averaging time. We assume the clock is driven on the $\ket{5/2,\pm1/2}\leftrightarrow\ket{3/2,\mp1/2}$ transition and use the calculated $\Delta f$ in Table~\ref{tab:my_label}. For our analysis, we assume $T_e=T_c=200$ s, and a SNR given by
\begin{equation*}
    S = \frac{N_{d}}{\sqrt{N_d + N_b}},
\end{equation*} where $N_b = b \times T_c$ is the total background counts, $b=6\times 10^4$ cps (based on the \mgf{} substrate background and scaling from a 20 nm to 100 nm film), and $N_d$ is the number of detected photons from the thorium isomer decay given by \begin{equation*}
    N_d = \eta N_e \left( 1 - e^{-\Gamma T_c}\right),
\end{equation*} where $N_e$ ($\sim2.7\times 10^{11}$) is the total number of \thor{} nuclei excited, $\eta$ = 0.01 is the assumed system detection efficiency, and $\Gamma$ is the transition decay rate. $N_e$ is calculated assuming that the probe laser linewidth is significantly smaller than the inhomogeneous Zeeman-limited transition linewidth, the probe laser power is \SI{1}{\micro W}, and all quenching mechanisms have been suppressed so that the maximum emitter fraction is achieved. From this we obtain a projected clock performance of $5 \times 10^{-17}/\sqrt{\tau}$.

\putbib[SI]
\end{bibunit}

\end{document}